\documentclass[prc,preprint]{revtex4}
\usepackage{graphicx,latexsym}
\usepackage{dcolumn}

\begin{document}

\title{Using the $\Delta_3$
statistic to test for missed levels in mixed sequence neutron
resonance data.}
\author{Declan Mulhall}
 \affiliation{Department of Physics/Engineering,
 University of Scranton, Scranton, Pennsylvania 18510-4642, USA.}
 \email{mulhalld2@scranton.edu}
\date{\today}
\begin{abstract}
The $\Delta_3(L)$ statistic is studied as a tool to detect
missing levels in the neutron resonance data where 2
sequences are present. These systems are problematic because
there is no level repulsion, and the resonances can be too
close to resolve. $\Delta_3(L)$ is a measure of the
fluctuations in the number of levels in an interval of length
$L$ on the energy axis. The method used is tested on
ensembles of mixed Gaussian Orthogonal Ensemble (GOE)
spectra, with a known fraction of levels ($x\%$) randomly
depleted, and can accurately return $x$. The accuracy of the
method as a function of spectrum size is established. The
method is used on neutron resonance data for 11 isotopes with
either s-wave neutrons on odd-A, or p-wave neutrons on
even-A. The method compares favorably with a maximum
likelihood method applied to the level spacing distribution.
Nuclear Data Ensembles were made from  20 isotopes in total,
and their $\Delta_3(L)$ statistic are discussed in the
context of Random Matrix Theory.
\end{abstract}


\maketitle

\section{\label{sec:intro}Introduction}
Neutron resonance data provides us with a list of eigenvalues
of the nuclear Hamiltonian. As such, they are a testing
ground for a range of  ideas in theoretical nuclear physics,
from level density models to quantum chaos
\cite{weid09,guhr}. The data sets are rarely complete, levels
are invariably missed, due to the finite resolution of the
experimental apparatus, or the weakness of the signal, or
some other factor. Whatever the reason for levels being
missed, it is important to have an estimate of how incomplete
a data set is. When the data is a mixture of 2 sequences of
levels (a sequence of levels is a set of levels with the same
quantum number), there is no level repulsion, levels can be
very close indeed, and the number of missed levels is
expected to increase. We can use Random Matrix Theory (RMT)
to estimate the number of missed levels. An analysis based on
RMT works because at the excitation energies involved,
$\approx 7$ MeV, the nucleus is a chaotic system and the
nuclear spectra have the same fluctuation properties as the
Gaussian Orthogonal Ensembles (GOE) \cite{brody, stock,
boh84}. Various statistics of RMT have been used to evaluate
the completeness of data in the past \cite{brody}. The most
popular statistic is the level spacing distribution, $P(s)$.
The $\Delta_3(L)$ statistic, introduced by Dyson \cite{dyson}
has traditionally been underexploited for this task. The
sensitivity of the $\Delta_3(L)$ statistic was reevaluated in
\cite{mulhall07}, where it was shown that the actual variance
of the $\Delta_3(L)$ statistic calculated from a specific
spectrum was smaller than the ensemble result. It was then
used for statistical spectroscopy to give information on
missed levels in a single sequence of pure neutron resonance
data. The situation is more complicated, however when the
data consists of 2 sequences of levels. In this case the
quantum number that differs between the sequences is spin.
When s-wave neutrons (angular momentum $L=0$) are incident on
an odd-A isotope, with spin-$j$, the neutron resonances have
two possible spin values, $j \pm \frac{1}{2}$. When p-wave
neutrons (angular momentum $L=1$) are incident on an
even-even (spin-zero) isotope, the possible spin values for
the resonances are $\frac{1}{2}$ or $\frac{3}{2}$. The level
repulsion that is present in single sequence data is gone,
and now levels can be very close indeed. This makes them
easier to miss by counting two separate levels as one, and
the number of missed levels may increase.

In this work the neutron resonance data of s-wave neutrons
incident on 7 odd-A isotopes, and p-wave neutrons on 4
even-even isotopes was analyzed with a RMT method, to gauge
the completeness of the data. The $\Delta_3(L)$ values for
each set of neutron resonance data were compared to the RMT
results from a numerical ensemble. There is no closed form
expression for the $\Delta_3(L)$ statistic for mixed,
depleted GOE spectra, so we found it numerically. A range of
ensembles of mixed depleted GOE spectra was made. Each
ensemble consisted of 500 spectra. Each spectra was a mixture
of 2 unfolded GOE spectra, mixed in a proportion
$\alpha:(1-\alpha)$, appropriate to the isotope in question.
Furthermore the spectrum size, $N$, was appropriate for
comparison with the experimental data. Each spectra in an
ensemble was depleted by $x$\%, by randomly deleting levels.
The probability of deletion was uniform across the spectrum
i.e. independent of energy. Other energy dependant
probability distributions were used, in an attempt to mimic
experimental resolution but the results were the same. There
were 21 ensembles for each value of $\alpha$ and $N$, one for
each value of $x$ from 0\% to 10\% in steps of 0.5\%. A
comparison of $\Delta_3(L)$ from the experimental spectrum
with the appropriate ensembles gave an estimate of $x$. These
results were compared with those of a MLM analysis based on
$P(s)$.

First, we make a few comments on the basic approach of using
random matrices to model real physical systems. Specific
nuclear energy levels are incalculable at neutron separation
energies. The system is too complex and the level density is
too large so that even a weak residual interaction would mix
whatever basis states you start with into a random
superposition. In other words, the Hamiltonian is, for all
practical purposes, random. Given a set of energy levels with
the same quantum numbers (a pure sequence) from a complex
system, in a region of high level density, the only clue
available to the system it came from would be the functional
form of the level density itself, this is  a so-called
secular variation. If you removed this system-specific
information by rescaling the energies so that the adjusted
spectrum had a uniform level density, and an average spacing
of one, then all that would remain of your original spectra
are the fluctuation properties. It turns out that these
fluctuation properties are rich in information about the
system. This rescaling is known as unfolding
\cite{guhr,brody}. It is a powerful idea, because it strips
away all the details that depend on the specifics of the
system. The only surviving features are in the statistical
fluctuations of the spectra, and these are due to the global
symmetries of the system. These fluctuations are the
variables of RMT. It is the purpose of this work to use one
of them, $\Delta_3(L)$, to estimate the number of levels
missed in neutron resonance experiments.

In the next section we describe the $\Delta_3(L)$ statistic
and give the GOE results for the cases of complete single and
mixed spectra. The distinction between spectral and ensemble
averages is clarified. The details of the calculation of GOE
spectra, the unfolding procedure, and the evaluation of
$\Delta_3(L)$ are explained. In Sect.~\ref{sec:method}, the
method of determining the fraction of levels missing from a
spectrum, using both $\Delta_3(L)$ and $P(s)$ is explained.
Both methods are tested on depleted, mixed GOE spectra.
Sect.~\ref{sec:nde} sets of neutron resonance date are
themselves grouped into Nuclear Data Ensembles, and a
qualitative comparison with the GOE is made. In
Sect.~\ref{sec:results} we describe the experimental data
sets and discuss the results.

\section{\label{sec:def}The $\Delta_3(L)$ statistic: definition and calculation}
The $\Delta_3(L)$ statistic is a robust statistic for a RMT
analysis, revealing a remarkable long range correlation in
chaotic spectra. It is a measure of the variance in the
number of levels in an interval of length $L$  anywhere on
the energy axis. It is defined in terms of the cumulative
level number, ${\mathcal N}(E)$, the number of levels with
energy less than or equal to $E$.  A graph of ${\mathcal
N}(E)$ is a staircase function, each step is one unit high,
and $s$ units deep, where $s$ is the level spacing. A
harmonic oscillator spectrum has no fluctuations, so in this
case ${\mathcal N}(E)$ will look like stairs at a
$45^{\circ}$ angle. In the case of a GOE spectrum, random
fluctuations will make ${\mathcal N}(E)$ deviate from the
regular staircase. The $\Delta_3(L)$ statistic measures this
deviation. It is defined by:
\begin{eqnarray}
\nonumber \Delta_{3}(L) &=&\left \langle {\rm min}_{A,B}\;
\frac{1}{L}\;\int^{E_i+L}_{E_i}dE'\,[\;{\mathcal
N}(E')-AE'-B]^{2}\;
\;\right\rangle \\
&=&\langle \Delta^i_3(L) \rangle\:, \label{eq:d3}
\end{eqnarray}
where we use angle brackets  to denote the {\sl spectral
average} of a quantity, in this case, the average is over all
values of $i$, the location of the window of $L$ levels
within the spectrum. $A$ and $B$ have values that minimize
$\Delta^i_3(L)$; they are recalculated for each value of $i$.
In the case of the perfectly rigid harmonic oscillator, the
triangles between the staircase ${\mathcal N}(E)$, and the
straight line, $AE-B$, with $A=1$, $B=0$, will give us
$\Delta_3(L)=1/12$. At the other extreme, a classically
regular system will lead to a quantum mechanical spectrum
with no level repulsion, the fluctuations are far greater,
and $\Delta_3(L)=L/15$. Such a spectrum is referred to as
Poissonian. The asymptotic RMT result for the Gaussian
Orthogonal Ensemble (GOE) is
\begin{eqnarray}
\nonumber \Delta_{3}(L) &=&\frac{1}{\pi^2}\,\left[\log(2\pi
L)+\gamma-\frac{5}{4}-\frac{\pi^2}{8}\right]\;
\\
&=&\frac{1}{\pi^2}(\log L-0.0678)\:, \label{eq:d3th}
\end{eqnarray}
with $\gamma$ being Euler's constant. The ensemble variance
of $\Delta_3(L)$ is $\sigma_e^2 = (0.110)^2$.

The difference between ensemble and spectral averaging is
important here, and we need to be very clear on what this
means. When there is risk of confusion, we will use an
overline for an ensemble average, as opposed to the angle
brackets used for spectral averages. If you take a single GOE
spectrum, and look at any window of $L$ levels, calculate the
quantity inside the angle brackets in Eq.~\ref{eq:d3}, ($i$
will be fixed at some random value), then you have a single
number, i.e. $\Delta^i_3(L)$. Eq.~\ref{eq:d3th} gives the
average value of this number evaluated for many different
spectra, and the variance of this number is 0.110,
independent of the value of $L$. This is an ensemble average.
The definition (\ref{eq:d3}) implies a spectral average, but
(\ref{eq:d3th}) is an ensemble average. The GOE result in Eq.
\ref{eq:d3} could be written $\overline{\Delta}_3(L)$, but
traditionally the overline is dropped, as RMT results
typically pertain to ensemble averages. The ambiguous
notation, while standard, is unfortunate.

The practical job of using the $\Delta_3(L)$ statistic to
detect missed levels depends on the variance of the
statistic. The value $\sigma_e^2 = 0.110^2$ is too large to
be useful, but this is the ensemble result. When analyzing a
specific spectrum, we take a spectral average. The variance
\textit{within} a spectrum of $\Delta^i_3(L)$ is much
smaller. The window of $L$ levels is positioned everywhere on
the spectrum and $\Delta^i_3(L)$ is calculated for all values
of $i$. The average of these numbers is what we report as the
$\Delta_3(L)$ for the spectrum. This will not be independent
of $L$, as there will be $N-L+1$ possible values for $i$, so
the corresponding variance will be smaller for smaller $L$.
Neighboring values of $i$ will correspond to overlapping
windows, giving correlated values of $\Delta^i_3(L)$. In
general, RMT results like Eq.~\ref{eq:d3th} are for the {\sl
ensemble average} of $\Delta^i_3(L)$, with $i$ fixed. When
all values of $i$ are included, the ensemble average becomes
the ensemble average of a  spectral average, and the variance
is smaller. Brody et al. \cite{brody} suggest that when
non-overlapping windows are used in the spectral averaging,
then the appropriate variance should be
$\sigma^2=\frac{\sigma_e^2}{p}$, were $p$ is the number of
non-overlapping windows, $p \approx \frac{N}{L}$. This gives
a variance of $\sigma^2=(0.11)^2 \frac{L}{N}$. This is called
the Poisson estimate and is verified in \cite{mulhall07}.
This is the value that should be used when comparing
$\Delta_3(L)$ from a specific spectrum with an ensemble
value.

Dyson \cite{dyson} derived an expression for $\Delta_3(L)$
for $m$ independent spectra, superimposed in proportions
$f_1,\, f_2,\,\dots f_m$. Letting $\Delta_{3m}(L)$ be the
spectral rigidity of the $m^{\textrm{th}}$ sub-spectrum, he
showed $\Delta_3(L)=\sum_{i=1}^m \Delta_{3m}(f_i L)$. If the
$m$ spectra are all from the GOE, the ensemble variance is
$\sigma_e^2 = (0.110\,m)^2$. The Poisson estimate then gives
$\sigma = 0.110\,m\,\sqrt{L/N}$. Based on this number
$\Delta_3(L)$ can be used to distinguish between spectra with
$m=1$ or $m=2$ independent sequences present. The statistic
is not sensitive to the actual mixing fractions. There is an
assumption that the proportions $f_1,\, f_2,\,\dots f_m$ are
independent of energy. This may not be the case in neutron
resonances, the fraction of intruder p-wave resonances may
grow with energy.

In \cite{mulhall07} there is an extensive discussion about
the calculation and use of $\Delta_3(L)$. The main points are
that for  a practical analysis of an experimental spectrum
with $N$ levels, $\Delta_3(L)$ is as sensitive and useful a
tool for detecting missed levels as the level spacing
distribution, and that one should use the Poisson estimate
for error bars when comparing $\Delta_3(L)$ taken from a real
spectrum with an RMT result. The rest of this section will
elaborate on the calculational details of realizing a random
matrix ensemble, the unfolding procedure, and the calculation
of $\Delta_3(L)$.

A GOE spectrum is generated by diagonalizing a matrix with
normally distributed matrix elements, $H_{ij}$, having
\begin{eqnarray}
\nonumber P(H_{i\neq j})=\frac{1}{\sqrt{2 \pi
\sigma^2}}\,e^{-\frac{H_{ij}^2}{2 \sigma^2}}, \quad
P(H_{ii})=\frac{1}{\sqrt{4 \pi
\sigma^2}}\,e^{-\frac{H_{ii}^2}{4 \sigma^2}}
\end{eqnarray}
for the off-diagonal and diagonal elements respectively, all
with $\sigma=1$.  Each of the matrices has an approximately
semicircular level density, with $\rho(E)=\sqrt{4N-E^2},$ for
$|E|\leq 2\sqrt{N},\;0$ otherwise (see Mehta \cite{mehta} for
a discussion of deviations).

These GOE spectra have a semicircular level density. Nuclear
spectra have level densities that increase exponentially. To
compare one with the other, we must remove these secular
variations by rescaling the spectra so that they have a level
density of unity. This process is called unfolding. The usual
recipe for unfolding the GOE spectrum was followed
\cite{guhr,brody}: first extract the cumulative level density
${\mathcal N}(E)$, which will be a staircase function, from
the raw spectrum, next fit it to a smooth function, $\xi(E)$,
either numerically or analytically, and finally, using this
function, the $j^{\textrm{th}}$ level of the unfolded
spectrum is simply $\xi(E_j)$.

Given a spectrum of size $N$, a mixed spectra would be made
as follows: take an unfolded GOE spectrum of length $\alpha
N$, and and rescale it by dividing by $\alpha$. This will
make the level density smaller. Take another GOE spectra, of
length $(1-\alpha)N$, and rescale  it by dividing by
$(1-\alpha)$. Join and sort the two spectra. The result is a
spectrum of $N$ levels with a uniform level density of unity.
To make a  spectra of $N$ levels, with mixing $\alpha$, and
depletion $x$, then start with a mixed spectrum $N/(1-x)$,
and randomly remove a fraction $x$.

Following \cite{mulhall07} $\Delta_3(L)$ can be calculated
exactly. Using ${\mathcal N}(E)=i, E_i \leq E < E_{i+1}$, in
Eq. (\ref{eq:d3}), and performing the integral between two
adjacent levels, we come to
\begin{eqnarray}
\nonumber\Delta_3^i(L)=
\frac{1}{L}\;\sum^{i+L-1}_{j=i}\int^{E_j+1}_{E_j}dE'\,(j-AE'-B)^{2}\\
\nonumber= \frac{1}{L}\times(C+VA^2+WA+XAB+YB+ZB^2),
\label{eq:d3iexplicit}
\end{eqnarray}
where  $C = \sum^{i+L-1}_{j=i}j^2(E_{j+1}-E_j),\; V =
\frac{1}{3}(E_{i+L}^3-E_i^3),\;  W =
\sum^{i+L-1}_{j=i}-j(E_{j+1}^2-E_j^2),\;  X =
(E_{i+L}^2-E_i^2),\;  Y =
\sum^{i+L-1}_{j=i}-2j(E_{j+1}-E_j),\;  Z = (E_{i+L}-E_i)$.
Furthermore, the constraints $\partial (\Delta_3^i)/\partial
A = 0$ and $\partial (\Delta_3^i)/\partial B = 0$, lead to
the following expressions for $A$ and $B$ that minimize
$\Delta_3^i(L)$:
\begin{eqnarray}
\nonumber A=\frac{XY-2WZ}{4VZ-X^2}, \quad \nonumber B=
\frac{WX-2VY} {4VZ-X^2}.
\end{eqnarray}

\section{\label{sec:method}Calculating ensembles and using $\Delta_3(L)$}

\begin{figure}
\includegraphics[width=.4\textheight]{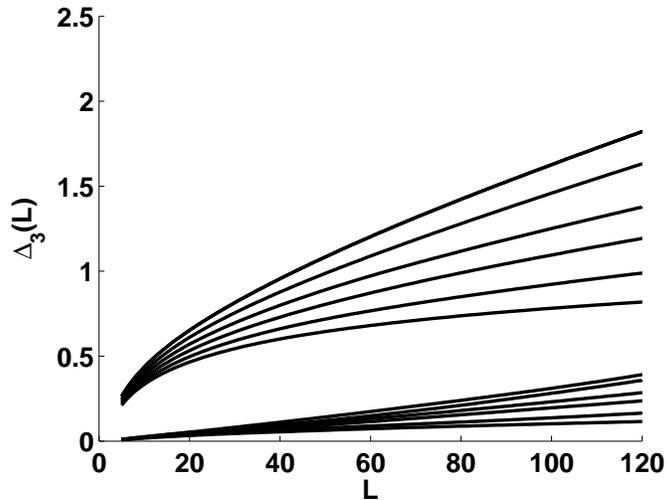}
\caption{\label{fig:dep}Here $\Delta_3(L,0.25,x)$ is shown
for $x=0,\,2,\,4,\dots 10$, (upper lines). The calculations
were for 500 spectra, with $N=400$.  The standard deviation,
$\sigma(400,L;x)$, corresponding to each value of $x$ (lower
lines) grows rapidly with $L$. This makes the statistic much
less sensitive for large $L$ values.}
\end{figure}

The task at hand is to compare the $\Delta_3(L)$ values from
an experimental data set with the GOE results for a single
spectrum consisting of 2 spectra mixed in proportion
$\alpha:(1-\alpha)$, and where $x\%$ of the levels have been
randomly depleted. We call this $\Delta_3(L,\alpha;x)$.
Ultimately, we will use the ensemble average,
$\overline{\Delta}_3(L,\alpha;x)$, of these depleted mixed
spectra to find the value of $x$ for an experimental data
set, with a specific $N$, and $\alpha$. There is a standard
deviation associated with this average which we write
$\sigma(N,L,\alpha;x)$. There is no analytical expression for
these quantities, so we will calculate them numerically, for
a range of values of $\alpha$, and $N$, and with $x$ ranging
from 0\% to 10\%. We chose an ensemble size of 500, and took
values of $N$ and $\alpha$ that allowed for a reasonable
comparison with the data. The dimension, $N$, refers here to
the number of levels {\it after} the fraction $x$ was
randomly removed. $\Delta_3(L)$ is calculated for each of
these spectra. The average of these 500 values is
$\overline{\Delta}_3(L,\alpha;x)$, the ensemble average, and
the standard deviation is $\sigma(N,L,\alpha;x)$. For each
value of $N$ and $\alpha$, there were 21 ensembles realized,
one for each value of $x$, which went from 0\% to 10\% in
0.5\% increments. The results for the ensemble with $N=400$,
and $\alpha=0.25$, are shown in Fig.~\ref{fig:dep}.

Now we have the tools necessary to compare a real spectrum
with depleted GOE spectra in a meaningful way. After
extracting $\Delta_3(L)$ from the data, the best value for
$x$ will be the one that minimizes
\begin{equation}
\chi^2(x) = \sum^{L_{\textrm{max}}}_{L_{\textrm{min}}}
\frac{[\Delta_3(L)-\overline{\Delta}_3(L,\alpha;x)]^2}{\sigma(N,L,\alpha;x)^2}.\label{eq:chi}
\end{equation}

\begin{table}
\caption{\label{tab:d3error}The fraction of depletion, $x$,
created; the fraction determined using the $\Delta_3(L)$
 method. The tests were run for 500 depleted spectra
of size $N=90$ and 400, and with $\alpha = 0.25$.  The mean
value, $\overline{x}$, and the standard deviation, $\sigma_x$
are given. } \label{tab:a}
\begin{ruledtabular}
\begin{tabular}{lll}
$x$ & $\overline{x} \,(N=90)$ & $\sigma_x$  \\
\hline
    3.2 \% & 3.2 \% & 3.1 \% \\
    5.3 \% & 4.7 \% & 3.4 \% \\
    8.2 \% & 7.3 \% & 2.9 \% \\
\hline
$x$ & $\overline{x} \,(N=400)$ & $\sigma_x$  \\
\hline
    2.9 \% & 3.0 \% & 1.8\%\\
    5.0 \% & 5.0 \% & 1.9\%\\
    8.1 \% & 7.9 \% & 1.6\%\\
\hline
\end{tabular}
\end{ruledtabular}
\end{table}

For practical purposes we need an estimate of the error in
$x$ from this method. To this end, we calculated the average
value of $x$, and its standard deviation, for 500 GOE spectra
with $N=90$ and 400 levels. The sets were made by randomly
deleting 3 levels from a spectra of 93 levels, 5 out of 95,
and 8 out of 98, to get spectra with $x= 3.23\%$, 5.26\%, and
8.16\% respectively. A similar was performed for $N=400$. The
results are in Table \ref{tab:d3error}. Our method gave good
agreement for the value of $x$, but for $N=90$ the
uncertainties in $x$, $\sigma_x$, were of the same order as
$x$, for example, with $x=5.3\%$, we get a value of $(4.7 \pm
3.4)\%$. However, for $N=400$, $\sigma_x$ dropped to 1.9\%
(close to a factor of $\frac{1}{2}$) which is to be expected.

While this process for estimating the uncertainties in the
fraction of missed levels is intuitive, and operationally
straightforward, the results may be optimistic. There are
more conservative approaches which would give larger
estimates of the uncertainties. A method of calculating
$\Delta_3(L)$ by Bohigas {\sl et al.} \cite{boh84} would have
us calculate $\Delta_3^i(L)$ for a much smaller number of
intervals, which overlap by $L/2$. In our scheme this would
translate into summing $i$ from 1 to $N-L-1$ in steps of
$L/2$ in Eq. \ref{eq:d3}. An estimate of the uncertainties
based on this scheme would give larger values, see
\cite{shriner92} for details.

The nearest level spacing distribution, $P(s)$, can be used
to test for missing levels also. Here we will follow the work
of Agvaanluvsan {\sl et al.} \cite{agv}, where the maximum
likelihood method is used to find the fraction of missing
levels in a sequence. We tested the method on mixed,
depleted, GOE spectra,  and compared the results with those
of our $\Delta_3$ analysis. $P(s)$ is known for a complete
mixed GOE spectrum \cite{guhr}. If a level is missing, then
two nearest neighbor spacings are unobserved, while one
next-to-nearest spacing is included as a nearest level
spacing, when it should not be. Furthermore, if $1-x$ is the
fraction of the spectrum that is observed, then $D_{{\rm
obs}}$, the experimental value for the average spacing, is
related to the true value by $D=(1-x)D_{{\rm obs}}$.
Agvaanluvsan {\sl et al.} show that
\begin{equation}
P(s)=\sum_{k=0}^{\infty} (1-x)x^kP(k;s),\label{eq:pos}
\end{equation}
where $P(k;s)$ is the distribution function for the
$k^{\textrm{th}}$ nearest neighbor spacing, $E_{k+i}-E_i$;
for $k=0$ this reduces to $P(0;s)=P(s)$.

Given a set of level spacings $\{s_i\}$, the likelihood
function is ${\mathcal L}=\prod_{i}P(s_i)$. We are after the
value of $x$ that maximizes ${\mathcal L}$ (although in
practice it is easier to work with $\ln({\mathcal
L})=\sum_{i}\ln P(s_i)$). The functions $P(k;s)$ in Eq.
\ref{eq:pos} are complicated to derive, so instead of a
closed form,  the functions were fitted to the empirical
distributions from the superposition of 1500 mixed GOE
spectra, each of length $N=2000$. This was performed for each
value of $\alpha$ relevant to the experimental data. Given
these functions we tested the method on depleted spectra.
Specifically, the procedure was tested on 1500 GOE spectra,
with $N=90$ and 400, with $\alpha=0.25$; and for $N=150$,
200, 250, and 400, with $\alpha = 0.4$. The depletion in each
case went from 0.0\% to 20\% in steps of 0.5\%.  The error in
$x$ from this method is taken to be the standard deviation of
the outputted values of $x$ for the 1500 input spectra with
known value of $x$. The results for $\alpha = 0.25$ are shown
in Table \ref{tab:mlmerror}. The agreement with the
$\Delta_3(L)$ method is encouraging. The $\Delta_3(L)$ method
seems to be slightly more accurate, but the uncertainty in
$x$ is slightly smaller for the MLM. Again, as with the
uncertainty estimates for the $\Delta_3(L)$ method, the MLM
uncertainties in $x$ could have been made a different way,
for example Agvaanluvsan {\sl et al.} \cite{agv}, use a
different criterion, based on the values of the likelihood
function. We have favored an approach based on distribution
of results of many simulations.

\begin{table}
\caption{\label{tab:mlmerror}The fraction of depletion, $x$,
created; the fraction determined using the maximum likelihood
method. The tests were run for 1500 depleted spectra of size
$N=90$ and 400, and with $\alpha = 0.25$.  The mean value
$\overline{x}$, and the standard deviation, $\sigma_x$ are
given. } \label{tab:b}
\begin{ruledtabular}
\begin{tabular}{lll}
$x$ & $\overline{x} \,(N=90)$ & $\sigma_x  $  \\
\hline
    3.0 \% & 4.1 \% & 3.1 \% \\
    5.0 \% & 6.1 \% & 3.4 \% \\
    8.0 \% & 9.7 \% & 2.8 \% \\
\hline
$x$ & $\overline{x} \,(N=400)$ & $\sigma_x  $  \\
\hline
    3.0 \% & 3.6 \% & 1.6\%\\
    5.0 \% & 5.6 \% & 1.6\%\\
    8.0 \% & 8.7 \% & 1.4\%\\
\hline
\end{tabular}
\end{ruledtabular}
\end{table}

\section{\label{sec:nde}Nuclear data ensemble}
\begin{figure}
\includegraphics[width=.4\textheight]{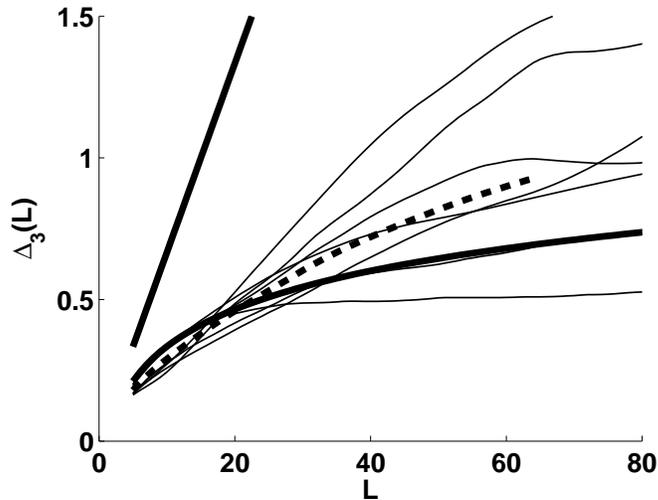}
\caption{\label{fig:d3oddA} $\Delta_3(L)$ for the 7 odd-A
isotopes of Tab.~\ref{tab:c} (thin lines). The dashed line is
the average of these thin lines. The lower thick line is the
GOE value for $\alpha=0.4$ with no depletion ($x=0\%$). The
upper thick line is for spectra with Poissonian statistics.
In cases where an isotope has 2 entries in Table \ref{tab:c}
the low energy subset is used here. }
\end{figure}

\begin{figure}
\includegraphics[width=.4\textheight]{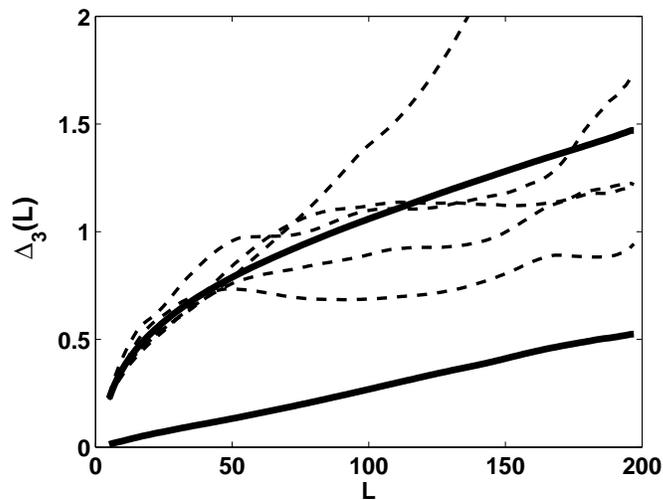}
\caption{\label{fig:d3rep}$\Delta_3(L)$, and its standard
deviation vs. $L$ for 5 randomly chosen GOE spectra (dashed
lines), with 4\% depletion, $\alpha = 0.33$, and $N=200$. The
upper solid line is the ensemble average. The lower solid
line is the corresponding $\sigma$. }
\end{figure}

The main idea behind using RMT to describe the fluctuations
in nuclear spectra, is that each isotope corresponds to a
random Hamiltonian from an ensemble with similar statistical
properties as the GOE. Neutron and proton resonance data sets
can be combined to make a so-called ``Nuclear Data Ensemble"
(NDE), see Haq \textsl{et al.} \cite{haq}. The odd-A isotopes
we analyzed can be regarded as a small NDE. In
Fig.~\ref{fig:d3oddA} we  have $\Delta_3(L)$ for each isotope
(light lines), and the ensemble average (dashed line). There
is a large variation in $\Delta_3(L)$ from spectrum to
spectrum, but compare this with the situation in
Fig.~\ref{fig:d3rep}, where  we see $\Delta_3(L)$ and
$\sigma(N,L,\alpha;x)$ for 5 randomly chosen samples taken
from an ensemble of GOE spectra, size $N=200$, with
$\alpha=0.33$ and $x=4\%$ depletion. It is clear that
variations in $\Delta_3(L)$ within the ensemble are normal.
This is reflected in the relatively large value of
$\sigma(N,L,\alpha;x)$. It is interesting to note that the
plots of the level spacing distribution do not have the same
variation: $P(s)$ for a particular spectrum will look like
the ensemble average, see \cite{mulhall07}.

\begin{figure}
\includegraphics[width=.4\textheight]{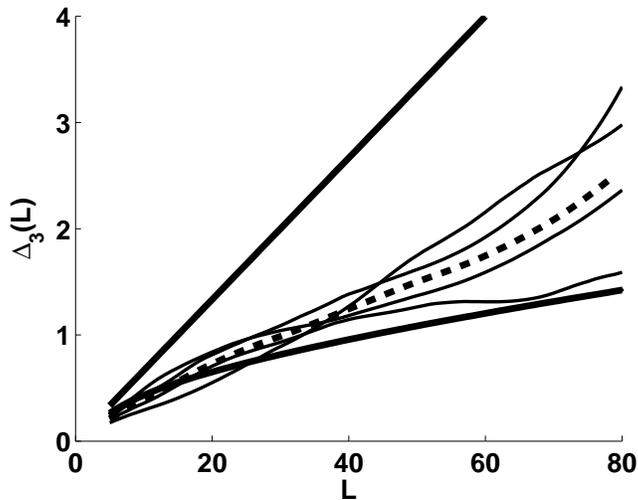}
\caption{\label{fig:d3L1} $\Delta_3(L)$ for the 4 data sets
of p-wave neutrons of Tab.~\ref{tab:d} (thin lines). The
dashed line is the average of these thin lines. The lower
thick line is mixed GOE spectra with $\alpha=0.4$ and 10\%
depletion. The upper thick line is for spectra with
Poissonian statistics. In cases where an isotope has 2
entries in Table \ref{tab:d} the low energy subset is used
here. }
\end{figure}

The ensemble plot made for an NDE consisting of the data from
p-wave neutrons is shown in Fig.~\ref{fig:d3L1}. The average
$\Delta_3(L)$ (dashed line) is consistent with an average of
$x > 10\%$. Compare it with the lower solid line which is
$\Delta_3(L)$ for an ensemble of GOE spectra with
$\alpha=0.33$ and $x=10\%$ depletion.

\begin{figure}
\includegraphics[width=.4\textheight]{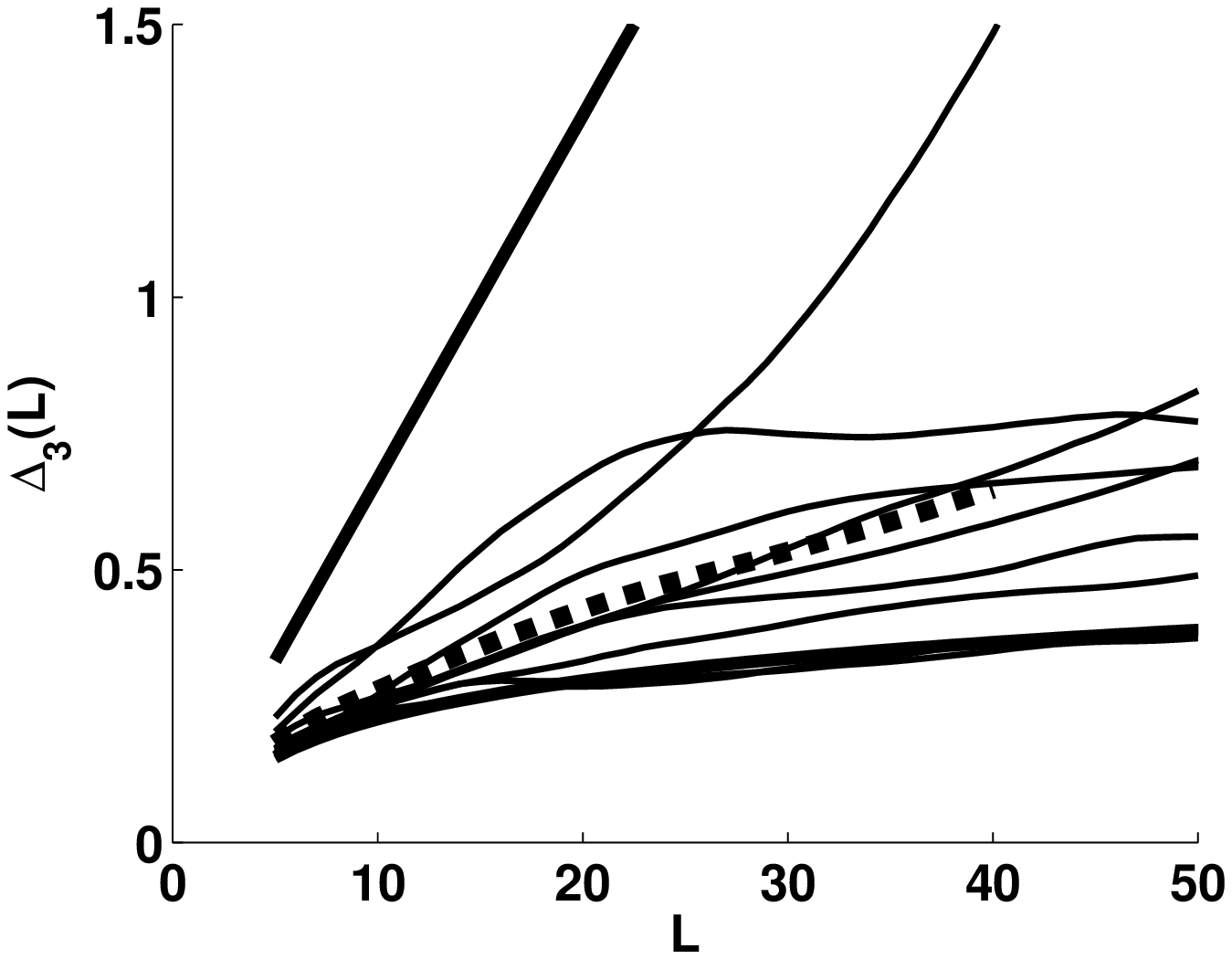}
\caption{\label{fig:d3evenA}Here we have $\Delta_3(L)$ for
the even-even NDE of the 9 isotopes described in
Ref.~\cite{mulhall07}, the dashed line being the average
value. Compare this with the pure GOE case (lower thick
line). The upper thick line is the Poissonian case.}
\end{figure}

In \cite{mulhall07} neutron resonance data from even-even
nuclei was analyzed. The ensemble plot corresponding to 9 of
these isotopes is shown in Fig.~\ref{fig:d3evenA}. The
isotopes represented here are $^{152}\text{Sm}$,
$^{152}\text{Gd}$, $^{154}\text{Gd}$, $^{158}\text{Gd}$,
$^{182}\text{W}$, $^{234}\text{U}$, $^{236}\text{U}$,
$^{242}\text{Pu}$, $^{240}\text{Pu}$. When put into
Eq.~\ref{eq:chi}, the average $\Delta_3(L)$ for this
even-even NDE gives a depletion of $x=4\%$. This is to be
taken lightly, as the data comes from different facilities,
over a period of decades, and is sometimes a combination of
data from different experiments. It does suggest that $x=4\%$
may be typical in neutron resonance experiments, when one
sequence of levels is present.

\section{\label{sec:results}Results and Discussion}
\begin{figure}
\includegraphics[width=.4\textheight]{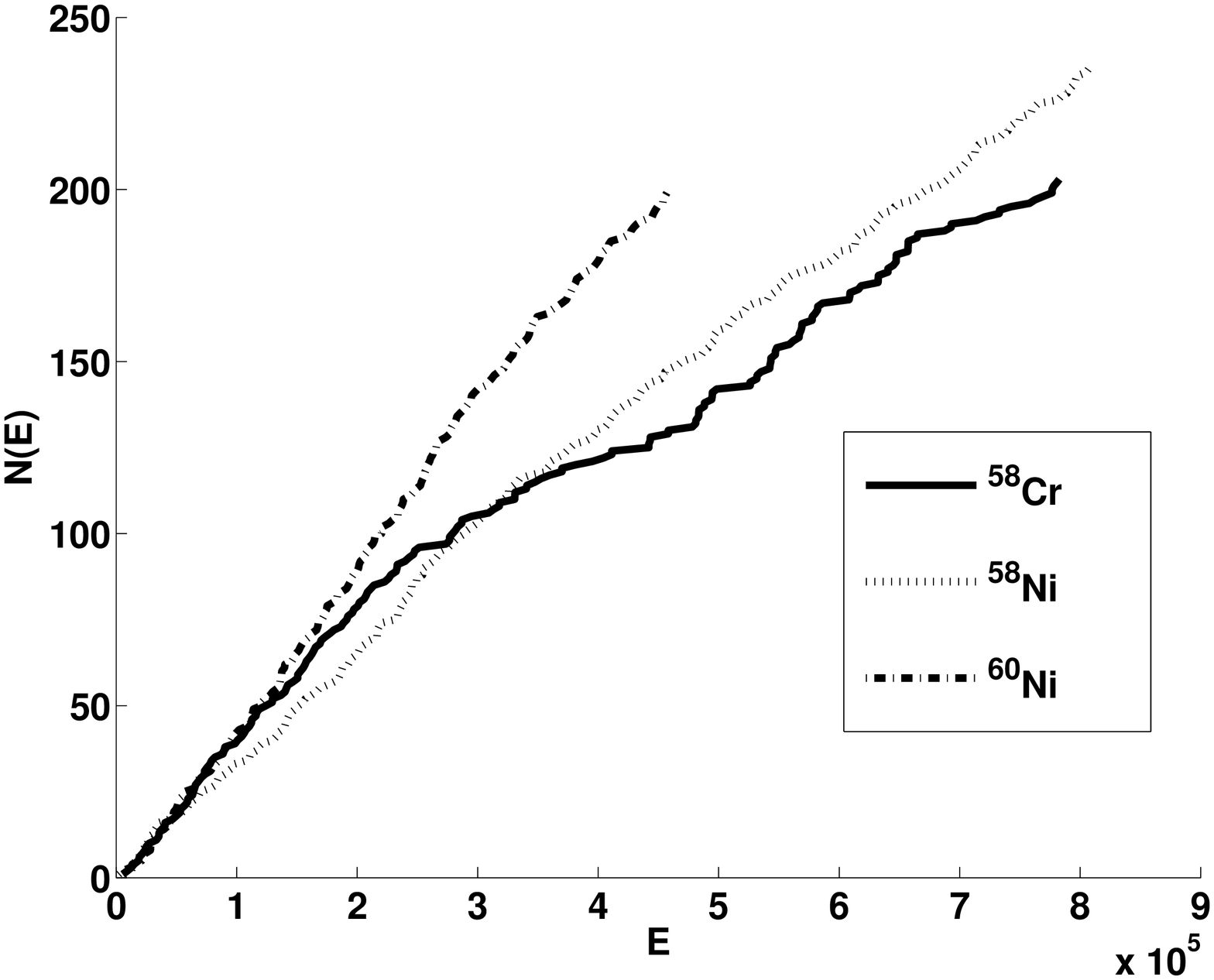}
\caption{\label{fig:NEcrni}${\mathcal N}(E)$ vs. $E$ for the
raw $^{58}\textmd{Cr}$, $^{58}\textmd{Ni}$,
$^{60}\textmd{Ni}$ data. Notice the decrease in slope after
the first 100 levels in the $^{58}\textmd{Cr}$ data,
suggesting that more levels were missed after the first 100
levels.}
\end{figure}

We analyzed neutron resonance data from 11 isotopes in all.
The data was taken from the  Los Alamos National Laboratory
website
\footnote{http://t2.lanl.gov/cgi-bin/nuclides/endind.}. The
s-wave resonances from 7 odd-A isotopes and the p-wave
resonances from 4 even-even isotopes were used. In the
even-even case, the spin labels were either 1/2 or 3/2, so
$\alpha = 1/3$ for all 4 sets. There was a variation in the
size of the sets. The ${\mathcal N}(E)$ plots gave us an
initial idea of the completeness of the data. In
Fig.~\ref{fig:NEcrni} we see ${\mathcal N}(E)$ for 3 raw data
sets of $^{58}\textmd{Cr}$, $^{58}\textmd{Ni}$ and
$^{60}\textmd{Ni}$ data. The range of energy is so small
compared to the neutron separation energy that one would
expect ${\mathcal N}(E)$ to have a constant slope. A kink,
where the slope (level density) gets suddenly smaller,
suggests an experimental artifact that leads to missed
levels.  This was quite common in the data sets, which
sometimes consisted of data from different facilities. In
Fig.~\ref{fig:NEu235}, for example, the see the case of
$^{235}\textmd{U}$. There are 3 kinks in ${\mathcal N}(E)$,
starting at $E=550$ eV. The histogram of the raw 3150 neutron
resonances for $^{235}\textmd{U}$ is shown in Fig.
\ref{fig:dos} with the corresponding discontinuities in the
level density. Notice that the kink at 550 eV in Fig.
\ref{fig:NEu235} corresponds to the obvious decrease in Fig.
\ref{fig:dos}. Another example is in the $^{50}\textrm{Cr}$
data, where we see a kink at $\mathcal{N}(E)=96$ in Fig.
\ref{fig:NEcrni}. When such a kink was observed, the lower
energy subset of the data (all the levels up to the kink) was
examined separately. After selecting data sets the spectra
were unfolded. The procedure is the same as that described in
Sect.~\ref{sec:def}, but ${\mathcal N}(E)$ was fit by a
straight line, as using a higher-order polynomial would be
unphysical.

The $\Delta_3(L)$ statistic is calculated from the data, and
used to the find the value of $x$ that minimizes the quantity
$\chi^2(x)$ in Eq.~\ref{eq:chi}. The MLM was used also, and
the results compared in Table \ref{tab:data} and
\ref{tab:dataL1}. In what follows we compare and discuss
these results, taking the odd-A isotope data first. The
typical error in the value of $x$ reported from either method
is 2\%, based on the results in Tables \ref{tab:d3error} and
\ref{tab:mlmerror}.

\subsection{$^{103}\textmd{Rh}$ and $^{147}\textmd{Sm}$}
\begin{figure}
\includegraphics[width=.4\textheight]{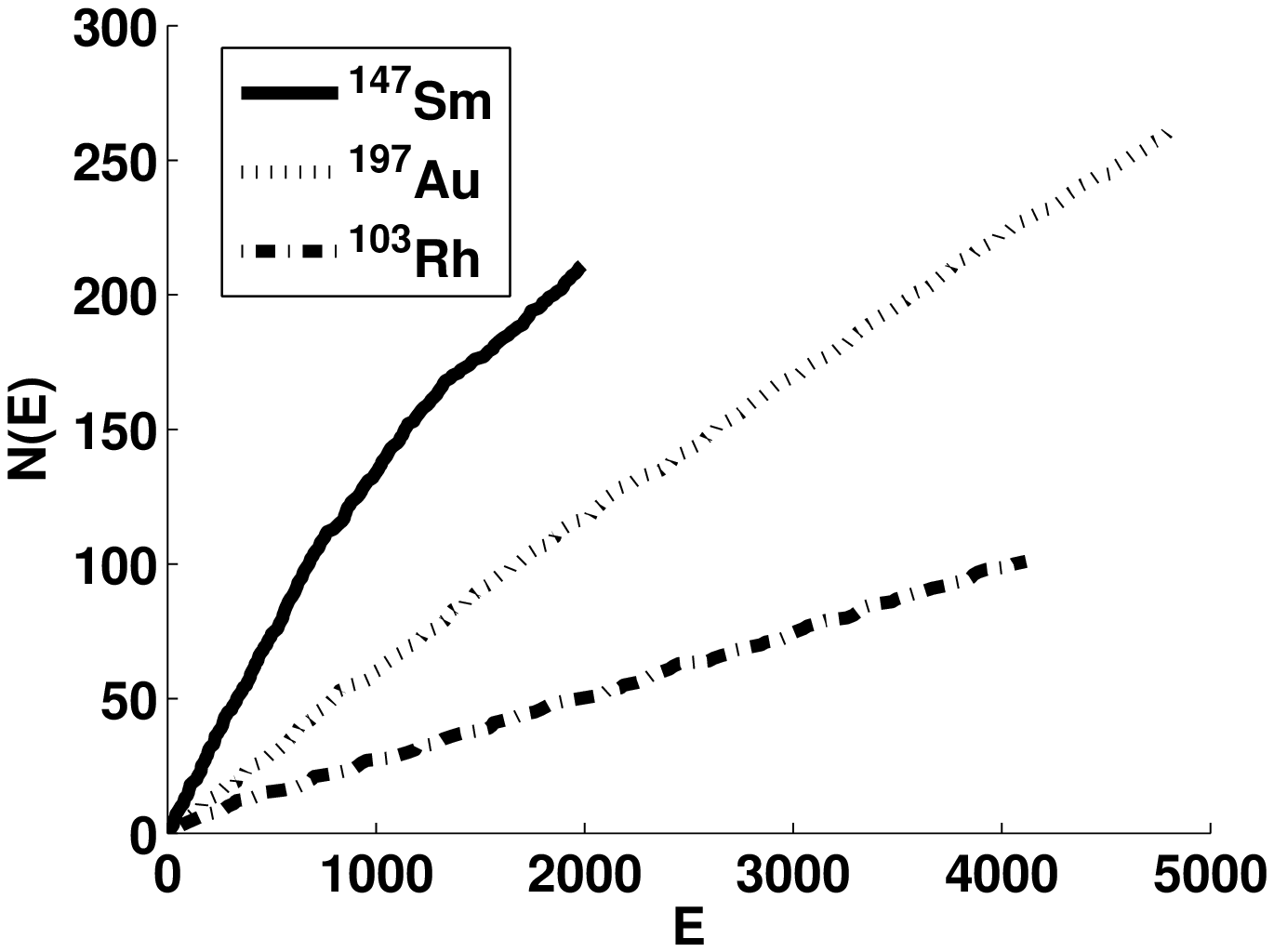}
\caption{\label{fig:NofESmAuRh}${\mathcal N}(E)$ vs. $E$ for
the $^{103}\textmd{Rh}$, $^{147}\textmd{Sm}$ and
$^{197}\textmd{Au}$ data. Notice the change in slope after
the $112^\textrm{th}$ level in the $^{147}\textmd{Sm}$ case.}
\end{figure}

These were the only two isotopes with consistent results.
Both methods suggested that only 2 or so levels were missed
from the 112 $^{103}\textmd{Rh}$ levels, while the first 112
levels of the $^{147}\textmd{Sm}$ data looked complete.
$\Delta_3(L)$ of $^{147}\textmd{Sm}$ lies on the $x=0\%$ line
of the ensemble result, see Fig.~\ref{fig:d3u235}. The
${\mathcal N}(E)$ plot for $^{147}\textmd{Sm}$ indicated
missed levels for the remainder of the spectrum, and this was
borne out by both methods, while the slope of ${\mathcal
N}(E)$ for $^{103}\textmd{Rh}$ was relatively constant. See
Fig.~\ref{fig:NofESmAuRh}. The $x=6\%$ from $\Delta_3(L)$ is
consistent with the MLM value of $x=9\%$, when reasonable
error bars of 2\% are used.

\subsection{$^{167}\textmd{Er}$}
The $^{167}\textmd{Er}$ data was missing 6\% of the levels
according to $\Delta_3(L)$, but the MLM said it was pure.
Looking at a plot of ${\mathcal N}(E)$ will not convince you
which value is preferred. There is no obvious kink, or
curvature.

\subsection{$^{185}\textmd{Re}$}
The agreement regarding the $^{185}\textmd{Re}$ data was fine
for the first 200 levels, both methods saying it was a
complete set. When the full set of 477 levels was examined
$\Delta_3(L)$ indicated over 10\% of the levels were missed,
while MLM still said 0\%. It is unlikely that there were 0\%
missed in the full set, as a slope of ${\mathcal N}(E)$
decreases with energy.

\subsection{$^{197}\textmd{Au}$}
The first 112 levels of $^{197}\textmd{Au}$ look incomplete
at the level of about 3\% according to both methods. When the
full set is analyzed, the MLM tells us that 7\% are missed,
but $\Delta_3(L)$ gives 0\%. This could be a normal
fluctuation, a reflection of the variation in the statistic
itself, as seen in Fig.~\ref{fig:d3rep}. An appeal to
${\mathcal N}(E)$ to support one result over the other is not
convincing, because the graph is straight. See Fig.
\ref{fig:NofESmAuRh}.

\subsection{$^{241}\textmd{Pu}$}
The results for $^{241}\textmd{Pu}$ are inconclusive. A
subset was made of the first 180 levels, based on ${\mathcal
N}(E)$. The last 57 levels had a slightly lower, but
constant, density. The MLM gave $x=0\%$ in both cases, while
$\Delta_3(L)$ gave $x=6\%$ in both cases. This disagreement
is similar to that found in $^{167}\textmd{Er}$ case.

\begin{figure}
\includegraphics[width=.4\textheight]{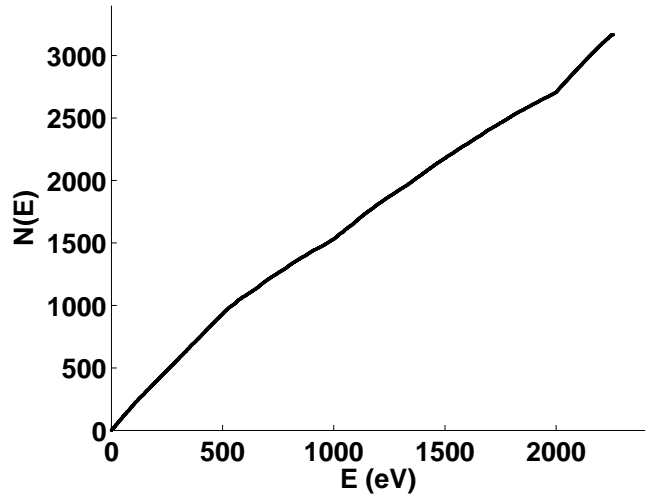}
\caption{\label{fig:NEu235}${\mathcal N}(E)$ vs. $E$ for the
raw $^{235}\textmd{U}$ data. Notice the decrease in slope
after the first 950 levels. There are 2 other kinks, at
$E=1000$ eV, and 2200 eV.}
\end{figure}

\begin{figure}
\includegraphics[width=.4\textheight]{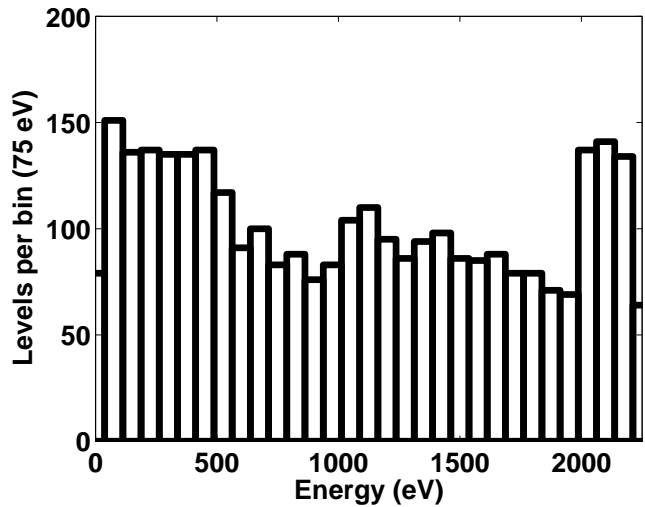}
\caption{\label{fig:dos}Histogram of the raw
$^{235}\textmd{U}$ data. This is the slope of
Fig.~\ref{fig:NEu235}. The binsize is 75 eV. The first 950
levels span the range $0<E<500$ eV. There is a drop in the
level density at 550 eV, and a sharp increase at 2000 eV,
where the level density nearly doubles. This is almost
certainly artificial, due to the fact that the data is the
union of data from different facilities.}
\end{figure}

\subsection{$^{235}\textmd{U}$}
\begin{figure}
\includegraphics[width=.3\textheight]{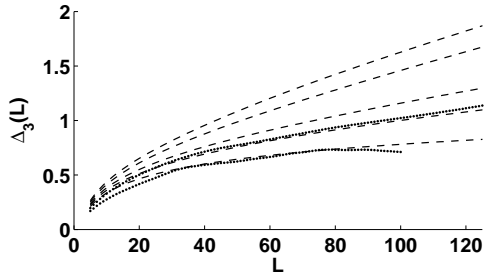}
\caption{\label{fig:d3u235}$\Delta_3(L)$ vs. $L$ for the
lowest 950 levels in the $^{235}\textmd{U}$ (upper dots), and
the lowest 112 levels of the $^{147}\textmd{Sm}$ (lower
dots). The dashed lines are the GOE values for a mixed
($\alpha = 0.4$) spectra, depleted by $x=0\%$, 3\%, 5\%, 8\%
and 10\%, starting from the lowest curve. The $\Delta_3(L)$
analysis suggests that there are no levels missed in the
$^{147}\textmd{Sm}$ data, and  3\% of the $^{235}\textmd{U}$
data missed.}
\end{figure}

The $^{235}\textmd{U}$ is an amazing data set, because of its
size and resolution. The first 950 levels were examined, as
suggested by the location of the first kink in
Fig.~\ref{fig:NEu235}, at an energy of 550 eV. The abrupt
change in the level density can also be seen in
Fig.~\ref{fig:dos}. The superposition of $j=3$ and 4 energies
were analyzed together and separately. The MLM gave
inconsistent results, saying the mixture of both sequences
was complete, but each individual sequence was missing 2\% or
3\%. $\Delta_3(L)$ gave 1.5\% missed in each separate subset.
It gave 3\% instead of the more consistent 1.5\% for the
superposition of both sequences, but these numbers agree with
each other within the bounds of error.  The errors in these
values of $x$ are about 1\%, given that $N=950$ and the
errors when $N=90$ are $\approx 3\%$. This value of $x= 3\%$
is very convincing when a graph of $\Delta_3(L)$ is examined,
see Fig.~\ref{fig:d3u235}.

\subsection{p-wave neutrons on even-even nuclei}
The $\Delta_3(L)$ analysis gave $x>10\%$ for all 4 isotopes,
while the MLM gave more credible results. In the case of
$^{58}\textmd{Ni}$, the decrease in the slope of ${\mathcal
N}(E)$ suggested that the first 116 levels looked like a more
complete spectrum than the full set of 236 levels,
Fig.~\ref{fig:NEcrni}. The result of the MLM analysis was
consistent with this, giving $x=5.3\% $ and $7.9\%$ for these
sets. The situation was the same for the $^{238}\textrm{U}$
data, with MLM giving $x=2.8\%$ for the first 300 levels, and
$x=5.4\%$ for the full set of 1130 levels. See Table
\ref{tab:dataL1}.

\begin{figure}
\includegraphics[width=.4\textheight]{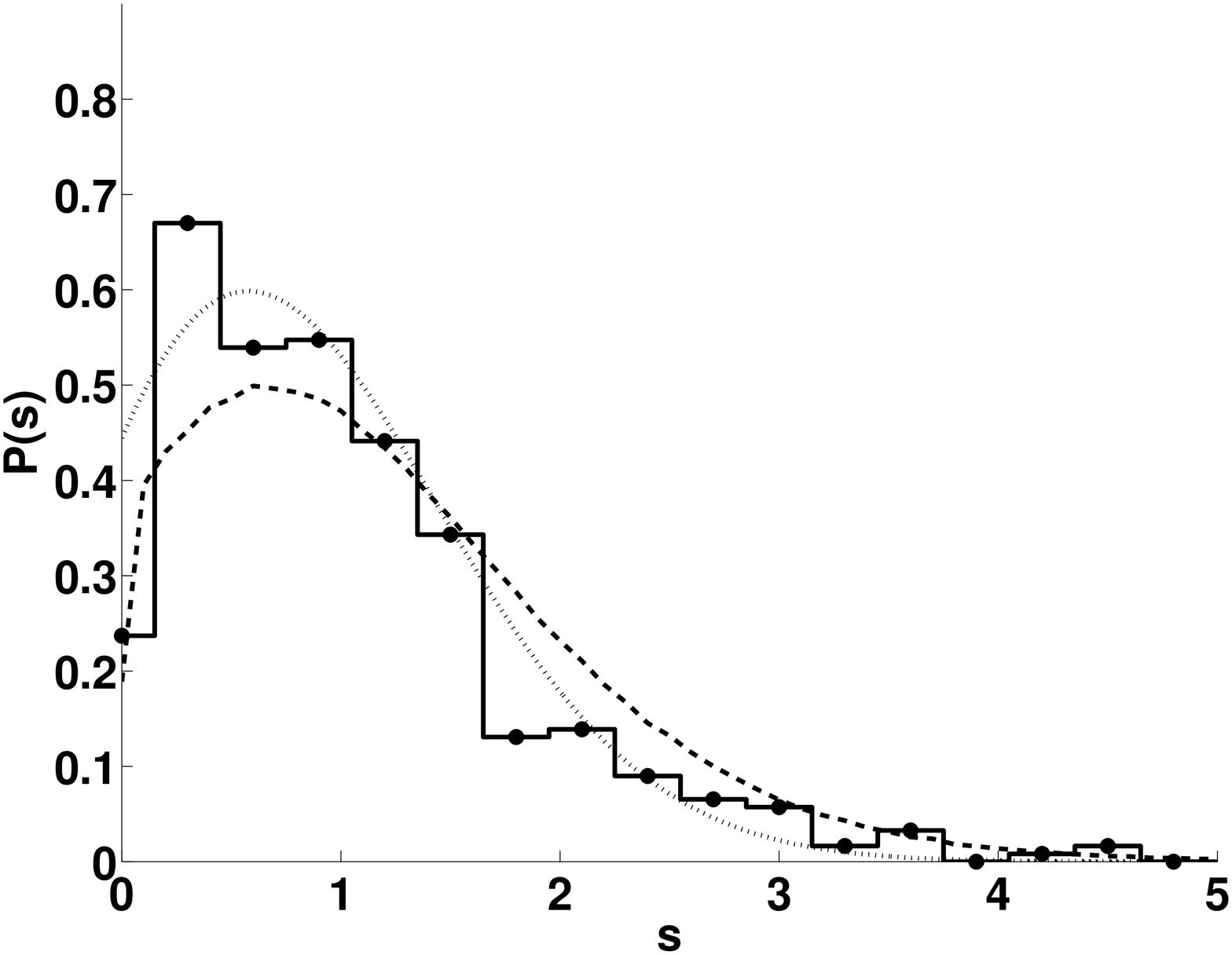}
\caption{\label{fig:posCrNi}$P(s)$ vs. $s$ for the combined
$^{50}\textmd{Cr}$ $^{58}\textmd{Ni}$ and $^{60}\textmd{Ni}$
data. The histogram has a bin width of 0.3, and has been
normalized so that its area is unity. Also shown are the GOE
results for $x=0\%$ (dotted line) and $x=10\%$ (dashed
line).}
\end{figure}

\begin{figure}
\includegraphics[width=.4\textheight]{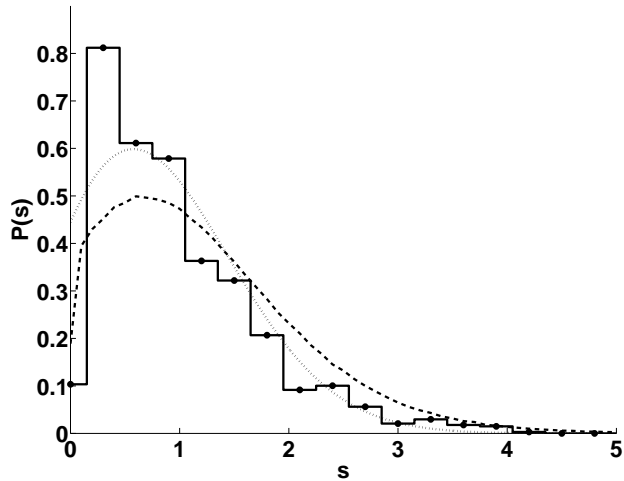}
\caption{\label{fig:posU}$P(s)$ vs. $s$ for the
$^{238}\textmd{U}$ data. As in Fig.~\ref{fig:posCrNi}, the
GOE results for $x=0\%$ (dotted line) and $x=10\%$ (dashed
line) are shown. }
\end{figure}

The large discrepancy in the values for $x$ is troublesome.
To shed more light on this histograms of the level spacings,
$s$ were compared with the GOE results for mixed spectra,
with $x=0\%$ and $x=10\%$ depletion. In this case, where the
possible labels for angular momentum are $1/2$ and $3/2$, we
used $\alpha = 1/3$. The lowest  96 $^{50}\textmd{Cr}$
levels, 116 $^{58}\textmd{Ni}$ levels, and the full set of
199 $^{60}\textmd{Ni}$ levels were combined into one set, see
Fig.~\ref{fig:posCrNi}, and all the $^{238}\textmd{U}$ levels
were considered as a separate set, see Fig.~\ref{fig:posU}.
We see in the combined set that only in the tail of $P(s)$ is
the data consistent with $x \approx 10\%$, otherwise it looks
undepleted. The situation for the 1130 $^{238}U$ levels is
also perplexing in that there is an excess of small ($s<1.2$)
spacings, while the tail of the histogram is consistent with
$x=0$. In both cases the pictures suggest that a
$\Delta_3(L)$ is appropriate here, and the $x>10\%$ is very
unlikely indeed. Such a high fraction of missed levels would
suggest that there would be a higher number of large spacings
then we are seeing.

\begin{table}
\caption{\label{tab:data}The results for $x$, the percent of
missing levels in the data. In the last 3 rows, we separated
the 950 levels from the $^{235}\textmd{U}$ data into the
$j=3$ and $j=4$ sequences, according to the spin labels
assigned in the data.} \label{tab:c}
\begin{ruledtabular}
\begin{tabular}{llllll}
Isotope & MLM & $\Delta_3(L)$ & $N$ (\# levels) & subset \\
\hline
$^{103}\textmd{Rh}$ & 1.6\% & 2\%  &   101& All \\
$^{147}\textmd{Sm}$ & 0.0\% & 0.0\%  &    211  & $1 \rightarrow 112$ \\
$^{147}\textmd{Sm}$ & 9.9\% & 6.0\%  &    211  & All \\
$^{167}\textmd{Er}$ & 0.0\% & 6.0\%  &    113  & All \\
$^{185}\textmd{Re}$ & 0.0\% & 0.5\%  &    477  & $1 \rightarrow 200$ \\
$^{185}\textmd{Re}$ & 0.9\% & $>$10\%  &    477  & All \\
$^{197}\textmd{Au}$ & 2.6\% & 4.0\%  &    262  &$1 \rightarrow 119$ \\
$^{197}\textmd{Au}$ & 7.0\% & 0.0\%  &    262  &$120 \rightarrow 262$ \\
$^{241}\textmd{Pu}$ & 0.0\% & 6.0\%  &    237  &$1 \rightarrow 180$ \\
$^{241}\textmd{Pu}$ & 0.0\% & 6.0\%  &    237  & All \\
$^{235}\textmd{U}\quad j=3,4$ & 0.0\% & 3.0\%   &    3168  & 950 \\
$^{235}\textmd{U}\quad j=3$ & 3.4\% & 1.5\%  &    1436  & 381 \\
$^{235}\textmd{U}\quad j=4$ & 2.1\% & 1.5\%  &    1732  & 569

\end{tabular}
\end{ruledtabular}
\end{table}

\begin{table}
\caption{\label{tab:dataL1}The even-even nuclei had
resonances with orbital angular momentum $L=1$, which means
that these resonances had angular momentum $j=\frac{1}{2}$ or
$j=\frac{3}{2}$, with mixing parameter $\alpha=\frac{1}{3}$.
In these cases the $\Delta_3$ statistic gave $x>10\%$. The
MLM was more optimistic. An uncertainty of 2\% is reasonable
for the MLM and $\Delta_3(L)$ results} \label{tab:d}
\begin{ruledtabular}
\begin{tabular}{llllll}
Isotope & MLM & $\Delta_3(L)$ & $N$ (\# levels) & subset \\
\hline
$^{50}\textmd{Cr}$ & 3.9\% & $>$10\% & 203 &  $1 \rightarrow 96$  \\
$^{58}\textmd{Ni}$ & 5.3\% & $>$10\% & 236 & $1 \rightarrow 116$ \\
$^{58}\textmd{Ni}$ & 7.9\% & $>$10\% & 236 & All \\
$^{60}\textmd{Ni}$ & 11.6\% & $>$10\% & 199 & All \\
$^{238}\textmd{U}$ & 2.8\% & $>$10\% & 1130 & $1 \rightarrow 300$ \\
$^{238}\textmd{U}$ & 5.4\% & $>$10\% & 1130 & All

\end{tabular}
\end{ruledtabular}
\end{table}

\section{\label{sec:data}Conclusion}
The $\Delta_3(L)$ statistic was used to determine the
completeness of neutron resonance data for 7 different
odd-$A$ isotopes, and p-wave neutrons on 4 even-even
isotopes. These are difficult data to work with due to the
presence of 2 independent sequences of levels that do not
repel each other. This means there is a much higher incidence
of small level spacing, making it much more difficult to get
a complete set of data. A method of estimating the fraction,
$x$ of missed levels, based on $\Delta_3(L)$ was presented.
The method was tested on numerical realizations of depleted
and mixed GOEs. Experimental data was grouped into Nuclear
Data Ensembles, and $\Delta_3(L)$ calculated. The behavior
was consistent with the overarching theme of RMT. Results
were compared with the maximum likelihood method. There were
13 data sets made from the 7 odd-A isotopes (including
subsets for the same isotope). The MLM was in agreement with
the $\Delta_3(L)$ method in 7 out of these 13 data sets. Of
the 6 sets where there was discordance, it looked like there
was only one case where the MLM result made more intuitive
sense, ($^{197}\textmd{Au}$, levels 120 to 262). In the other
5, it is difficult to say which method, if any, is more
likely to be correct. In the 6 data sets made from the 4
even-even isotopes, the MLM gave a variety of consistent
results, while $\Delta_3(L)$ gave $x > 10\%$ for all sets. A
plot of $P(s)$ and a comparison with the GOE results suggest
that a $\Delta_3(L)$ analysis is not appropriate here. The
cumulative level number ${\mathcal N}(E)$ was used as another
indicator of the purity of the data, and the credibility of
the results from the two statistics. A strong case has been
made for the usefulness of the $\Delta_3(L)$ statistic as a
gauge of the completeness of a data set when a RMT analysis
is appropriate.
\begin{acknowledgments}

We wish to acknowledge the support of the Office of Research
Services of the University of Scranton,  and M. Moelter for
many fruitful discussions. Also we are grateful to the
anonymous referee who suggested the level spacing plots for
the $p$-wave resonances.

\end{acknowledgments}

\bibliography{d3neutronresPRC}

\end{document}